\documentclass[10pt,twocolumn,letterpaper]{article}

\usepackage{cvpr}      
\usepackage{amsmath}
\usepackage{algorithm}
\usepackage{algpseudocode}
\usepackage{graphicx} 
\usepackage{amssymb} 
\usepackage{amsfonts} 
\usepackage{algorithm, algpseudocode} 
\usepackage{caption} 
\usepackage[accsupp]{axessibility}  

\makeatletter
\let\@LN@col\relax
\let\@LN\relax
\makeatother

\def\confName{CVPR}
\def\confYear{2025}

\title{ATP: Adaptive Threshold Pruning for Efficient Data Encoding in Quantum Neural Networks}

\author{
Mohamed Afane\textsuperscript{1, *} \quad
Gabrielle Ebbrecht\textsuperscript{1} \quad
Ying Wang\textsuperscript{2} \quad
Juntao Chen\textsuperscript{1, *} \quad
Junaid Farooq\textsuperscript{3}\\
\textsuperscript{1}Fordham University \quad
\textsuperscript{2}Stevens Institute of Technology \quad
\textsuperscript{3}University of Michigan-Dearborn
}

\begin{document}
\maketitle

\renewcommand{\thefootnote}{*}
\footnotetext{Corresponding authors: \texttt{\{mafane,jchen504\}@fordham.edu}}

\begin{abstract}
Quantum Neural Networks (QNNs) offer promising capabilities for complex data tasks, but are often constrained by limited qubit resources and high entanglement, which can hinder scalability and efficiency. In this paper, we introduce Adaptive Threshold Pruning (ATP), an encoding method that reduces entanglement and optimizes data complexity for efficient computations in QNNs. ATP dynamically prunes non-essential features in the data based on adaptive thresholds, effectively reducing quantum circuit requirements while preserving high performance. Extensive experiments across multiple datasets demonstrate that ATP reduces entanglement entropy and improves adversarial robustness when combined with adversarial training methods like FGSM. Our results highlight ATP’s ability to balance computational efficiency and model resilience, achieving significant performance improvements with fewer resources, which will help make QNNs more feasible in practical, resource-constrained settings.
\end{abstract}

\section{Introduction}
Quantum machine learning (QML) has gained attention for its ability to solve problems that are difficult for classical models by using the unique properties of quantum systems \cite{abbas2021power}. It has shown practical benefits in fields such as chemistry \cite{sajjan2022quantum}, optimization \cite{hadfield2019quantum}, and others involving high-dimensional or structured data \cite{tian2023recent, biamonte2017quantum}.However, quantum algorithms are still constrained by the availability of qubits, noise, and decoherence in hardware \cite{preskill2018quantum, knill2005quantum}. Because of quantum hardware’s sensitivity to the environment, qubits are prone to premature collapse that degrades the fidelity of computation. Qubits may also be disturbed by signals intended to alter the state of another qubit in close proximity \cite{ash2020experimental} which reduces stability in hardware with many qubits and complicates the training of quantum models. As a result, shallower Quantum Neural Networks (QNNs) are often more practical. \cite{cerezo2022challenges, wang2024shallow}.

\begin{figure}[h!]
\centering
\includegraphics[width=0.48\textwidth, height=0.25\textheight]{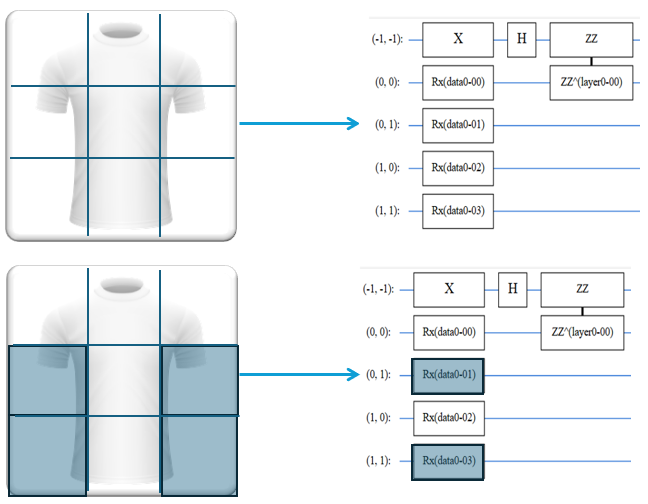}
\caption{Demonstration of the adaptive threshold pruning framework. The original image (top) is split into a 3x3 grid, with each section assessed for information density. The pruned image (bottom) shows filtered regions that fall below a defined intensity threshold in blue indicating areas that do not contribute significantly to the classification task, effectively freeing qubits and optimizing resources by focusing only on key areas of high relevance.}
\label{fig:shirt_grid}
\end{figure}

Efficient data encoding into quantum states remains a crucial bottleneck for achieving scalability and accuracy in QML applications \cite{ranga2024quantum, weigold2020data}. For instance, limiting qubit interactions can reduce crosstalk and enhance stability, which is essential for noise resilience \cite{alam2022qnet}. While using larger and deeper QNNs can enhance a model’s ability to capture complex data relationships and potentially improve performance, increasing encoding depth introduces new challenges. As the depth of the encoding circuit grows, encoded quantum states tend to converge towards a maximally mixed state at an exponential rate, becoming increasingly indistinguishable from random noise \cite{li2022concentration}. This concentration effect leads to a lack of state diversity and the emergence of barren plateaus, where gradients vanish, significantly hindering the model’s learning process \cite{ortiz2021entanglement}. These effects are especially problematic when encoding is not adapted to the structure or scale of the data. Thus, QNNs must balance expressibility with computational demands and sensitivity to barren plateaus \cite{nguyen2022evaluation}.

\begin{figure}[h!]
\centering
\includegraphics[width=0.48
\textwidth, height=0.26\textheight]{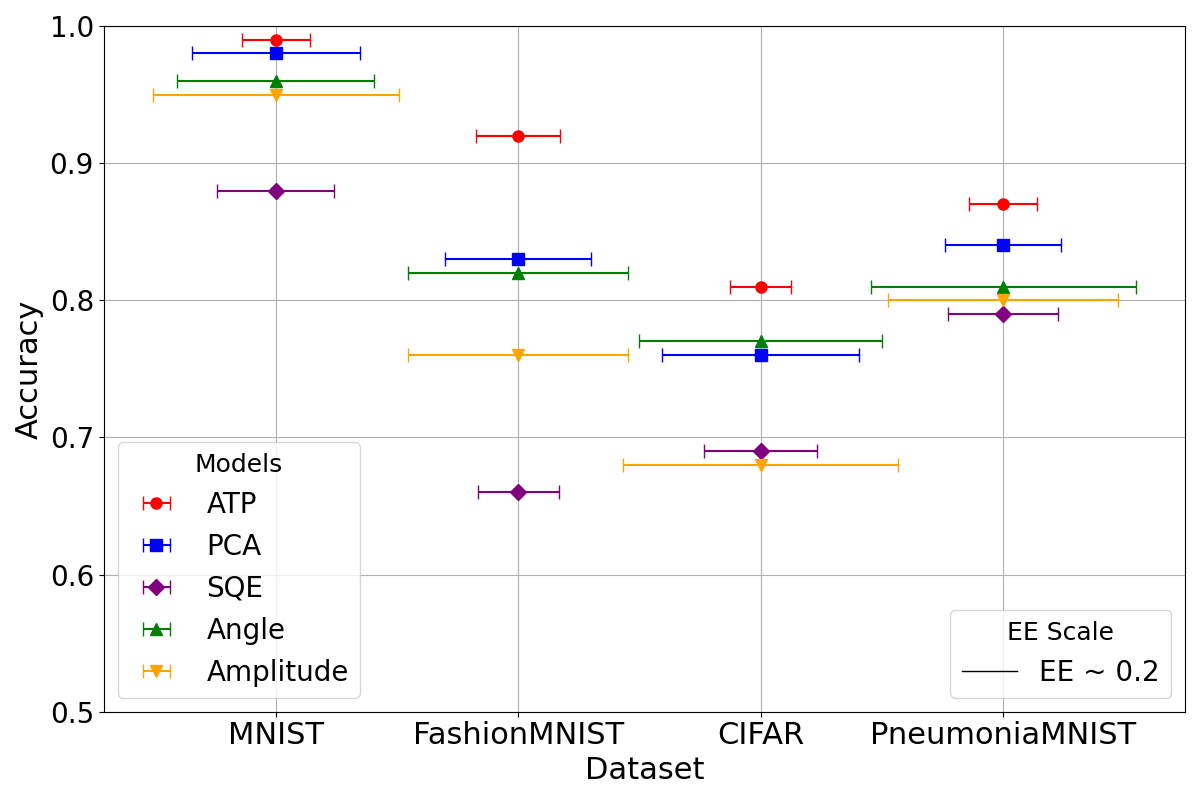}
\caption{Average accuracy across four datasets, comparing the performance of various encoding methods: Adaptive Threshold Pruning (ATP), Principal Component Analysis (PCA), Single Qubit Encoding (SQE), Angle, and Amplitude encoding. Horizontal error bars represent entanglement entropy (EE), with longer bars indicating higher entanglement. ATP generally achieves the highest accuracy with lower EE.}
\label{fig:acc_EE_comparison}
\end{figure}

While efficient data encoding can improve performance in classical models, such as by reducing training time or resource use, it is far more critical in QML. In quantum models, encoding directly impacts whether meaningful patterns can be learned in the first place, since poor encoding can waste limited qubit resources or amplify noise to the point that the model fails entirely \cite{lisnichenko2023quantum}. For instance, some methods only encode states with significant contributions to the model, allowing qubit resources to focus on meaningful data and enabling more effective scaling \cite{shee2022qubit}. Amplitude encoding is one such approach, representing data compactly to reduce the number of required qubits, though it limits the types of data and operations that can be used \cite{schuld2018supervised}. Other strategies, such as Qubit Lattice, a direct encoding method with low circuit depth, and FRQI, a compact scheme that uses fewer qubits, demonstrate different tradeoffs between qubit efficiency and processing flexibility \cite{lisnichenko2023quantum}. In general, encoding strategies that reduce resource overhead can optimize computation and limit potential errors by requiring fewer qubits to operate.

In this context, entanglement entropy (EE) serves as a key metric for assessing a model’s encoding efficiency. EE measures the degree of correlation between different qubits in a quantum system by quantifying the information shared between subsystems, typically calculated as the von Neumann entropy of a reduced density matrix of one part of a bipartite system. Higher EE indicates a greater level of interdependence between qubits, which can be advantageous for capturing complex data structures, but excessive entanglement may increase computational complexity and lead to overfitting. Anagiannis and Cheng, for instance, found that as EE rises in q-convolutional neural networks, the model’s cost function decreases, suggesting a link between structured entanglement and efficient learning \cite{anagiannis2021entangled}. Similarly, Martyn et al. demonstrated that while certain levels of entanglement improve training accuracy in quantum models, over-entangling the qubits can lead to diminished generalization and higher resource demands \cite{martyn2020entanglement}. Thus, EE not only provides insight into the information distribution across qubits but also helps balance model complexity and computational efficiency in quantum learning. 

To address these challenges, we introduce Adaptive Threshold Pruning (ATP), a technique that optimizes the encoding process by adaptively pruning non-essential features, thereby reducing both qubit usage and entanglement in the circuit, as visualized by Figure \ref{fig:shirt_grid}. ATP sets dynamic thresholds for feature pruning, selectively removing low-variance data to maintain model performance while lowering EE. Figure~\ref{fig:acc_EE_comparison} summarizes ATP’s performance compared to other encoding methods, demonstrating improvements in both accuracy and entanglement efficiency.

\section{Related Works}

\textbf{Masking:} In classical machine learning, masking strategies are commonly used in self-supervised learning by training models to reconstruct missing parts of the input from partial observations. Masked autoencoding, as used in the Masked Autoencoders (MAE) framework, achieves this by randomly hiding image patches and requiring the model to reconstruct them, encouraging the learning of meaningful representations without labels \cite{he2022masked}. Although the primary objective is representation learning, MAE improves efficiency by processing only visible patches through a lightweight encoder-decoder setup. This reduction in input complexity helps scale models to large datasets with fewer computational demands. Similar to how MAE filters input through random masking, our ATP approach uses threshold-based pruning to discard low-variance regions before encoding, allowing quantum models to focus on the most relevant parts of the data while reducing resource usage \cite{wang2024revisiting}.

\noindent\textbf{Pruning:} In contrast, pruning in QML has primarily been applied at the circuit level to reduce model complexity, addressing challenges such as noise and limited quantum resources. These techniques streamline QNNs by selectively removing parameters with minimal impact on accuracy, thereby enhancing computational efficiency and scalability. Hu et al. \cite{hu2022quantum} propose an architecture compression method for quantum circuits, achieving reduced circuit depth with minimal accuracy loss, while Wang et al. \cite{wang2022symmetric} demonstrate how pruning supports scalability in quantum neural networks. Circuit-level pruning methods, such as those by Sim et al. \cite{sim2021adaptive} and Kulshrestha et al. \cite{kulshrestha2024qadaprune}, adaptively eliminate non-essential parameters, optimizing quantum resources without compromising model performance. While architecture pruning addresses circuit complexity, efficient data encoding remains critical in quantum contexts, directly impacting qubit usage and entanglement management.

\noindent\textbf{Data encoding:} Several encoding methods have been developed to optimize qubit usage while preserving essential data features in quantum neural networks. For example, Single-Qubit Encoding (SQE) \cite{easom2022efficient} minimizes resource requirements by encoding data into a single qubit rather than multiple qubits. This technique uses a series of Z-Y-Z axis rotations on the Bloch sphere, parameterized by input data, to capture spatial relationships with minimal qubit and parameter usage. Classical methods like Principal Component Analysis (PCA) \cite{joliffe1992principal} and autoencoding \cite{goodfellow2016deep} also improve quantum model efficiency by reducing data dimensionality while retaining key features \cite{hur2022quantum}. PCA identifies principal components with the highest variance, focusing on informative aspects of the data, while autoencoders learn compact representations through non-linear dimensionality reduction. 

The choice of encoding strategy directly impacts the effectiveness of quantum methods by influencing qubit efficiency and circuit complexity. Angle encoding, commonly used for classification tasks, maps classical data to quantum gate rotations due to its simplicity \cite{rath2024quantum}. Amplitude encoding, which embeds data as probability amplitudes, allows quantum representation of exponentially large datasets but presents scalability challenges. Basis encoding represents binary data as computational basis states, suitable for binary data but less practical for complex datasets. Hybrid methods, such as amplitude-angle encoding, aim to combine strengths of individual techniques but often increase circuit depth, posing issues on noisy hardware with limited coherence times \cite{bhabhatsatam2023hybrid}. These limitations emphasize the need for adaptable encoding methods like ATP that optimize encoding based on data relevance and variability.

\noindent\textbf{Our approach:} ATP distinguishes itself by dynamically pruning  with adaptive thresholds. Unlike PCA, which applies linear transformations to capture data variance, ATP zeroes out less informative data points based on their relevance, adapting to each dataset’s unique structure. Unlike SQE, which minimizes resources by encoding data into a single qubit, ATP selectively retains expressive data features with minimal resource overhead. By concentrating quantum resources on essential data, ATP balances computational efficiency and entanglement requirements, offering a scalable solution tailored to the specific constraints and advantages of quantum systems.

To our knowledge, no prior work has addressed adaptive feature pruning within quantum data encoding. ATP thus fills this gap by offering a systematic approach to reduce quantum overhead while preserving model accuracy.

\section{Preliminaries}

\subsection{Quantum Computing Fundamentals}
Quantum computing exploits principles of quantum mechanics, providing a framework that differs profoundly from classical systems. In classical computing, data is represented by bits in distinct states of 0 or 1. Quantum computing, however, relies on \textit{qubits}, which can exist in linear combinations of 0 and 1 states, represented as:
\begin{equation}
|\psi\rangle = \alpha|0\rangle + \beta|1\rangle,
\end{equation}
where \( \alpha \) and \( \beta \) are complex numbers satisfying \( |\alpha|^2 + |\beta|^2 = 1 \). This property, known as \textit{superposition}, allows qubits to perform concurrent calculations, making quantum algorithms inherently parallel \cite{shor1998quantum, hirvensalo2013quantum}.

\subsubsection{Entanglement and Measurement}
\textit{Entanglement} is a quantum phenomenon in which qubits become correlated in such a way that the state of one qubit directly influences the state of another, regardless of physical distance \cite{preskill2012quantum}. For two entangled qubits, the state can be expressed as:
\begin{equation}
|\psi\rangle = \frac{1}{\sqrt{2}} \left( |00\rangle + |11\rangle \right),
\end{equation}
forming a maximally entangled Bell state. Entanglement is a crucial resource in QNNs, as it enables intricate data encoding and interaction patterns that cannot be replicated classically. However, the control and preservation of entanglement are challenging due to decoherence and noise, making efficient encoding schemes essential for managing entanglement and optimizing qubit usage \cite{gupta2020quantum}.

\subsection{Quantum Neural Networks}
QNNs use quantum circuits with parameterized gates, known as PQCs, to process data. A QNN layer can be defined by a series of parameterized unitary transformations \cite{zhao2021review}:
\begin{equation}
U(\theta) = \prod_{j} U_j(\theta_j),
\end{equation}
where \( \theta = \{\theta_j\} \) are tunable parameters. Each \( U_j \) represents a quantum gate that rotates or entangles qubits based on input data. These rotations enable the QNN to encode input features, transforming them into high-dimensional Hilbert space representations. During training, gradients of these parameters are optimized to minimize a loss function, analogous to classical neural networks, but constrained by quantum hardware limitations and noise sensitivity \cite{abbas2021power}.

\subsection{Data Encoding in Quantum Neural Networks}

Efficient data encoding is critical for the practicality of QNNs. In this work, we apply ATP before encoding data into quantum states. Primarily, we use \textit{angle encoding}, where each pixel value \( x_{i,j} \) of an image is mapped onto a rotation angle \( \theta_{i,j} \) for the RX gate:
\begin{equation}
\theta_{i,j} = \pi \cdot x_{i,j},
\end{equation}
resulting in the transformation \( RX(\theta_{i,j}) \) applied to each qubit \cite{ovalle2023quantum}. This method straightforwardly represents pixel intensities but can introduce redundancy for higher-dimensional data, as unnecessary values in qubit states increase entanglement and computational complexity.

We also conduct experiments using \textit{amplitude encoding}, which encodes normalized data values as amplitudes of a quantum state. For an input vector \( \mathbf{x} \) with dimension \(2^n\) (for \( n \) qubits), amplitude encoding transforms it to:
\begin{equation}
|\psi\rangle = \sum_{i=0}^{2^n-1} x_i |i\rangle,
\end{equation}
allowing all features to be encoded simultaneously in a single quantum state. While amplitude encoding can be efficient for certain data dimensions, it faces scalability challenges for large datasets.

Additionally, alternative encoding strategies exist, such as dense encoding techniques that represent multiple data features using fewer qubits by encoding them into different parameters of multi-qubit gates. These methods aim to balance representational power with resource constraints, but each encoding approach carries unique trade-offs in complexity and scalability. Across these various encoding schemes, ATP helps reduce redundancy by pruning non-essential features before encoding, leading to more efficient and robust QNN performance.

\subsection{Entanglement Entropy}

EE is a metric from quantum information theory that quantifies the correlation between different parts of a quantum system, typically measured using the von Neumann entropy of a reduced density matrix \cite{anagiannis2021entangled}. In QML, EE reflects a model's capacity to capture intricate data relationships. Higher EE often indicates stronger interdependence between qubits, which can increase the model's expressive power but also adds computational complexity and risks overfitting \cite{bai2022unsupervised}. Conversely, lower EE may represent efficient resource usage and reduced complexity, which is advantageous for simpler datasets \cite{liu2021entanglement}. Balancing EE is thus crucial, as it influences both learning potential and robustness in noisy environments. Therefore, achieving high performance with lower EE, or reduced complexity, is an important aim of this study.

\section{Methods}

To select effective threshold values for data pruning, we formalize threshold selection as a constrained optimization problem tailored to adapt to each dataset’s unique characteristics. This approach employs the Limited-memory Broyden-Fletcher-Goldfarb-Shanno algorithm with box constraints (L-BFGS-B), a quasi-Newton method that efficiently handles high-dimensional optimization by leveraging gradient-based updates to refine threshold levels. By using this optimization framework, we systematically balance accuracy with computational efficiency, ensuring that thresholds are adjusted to retain essential information while minimizing redundancy. The following sections present the mathematical formulation and algorithmic steps that drive this threshold-tuning process for enhanced QNN performance.

\subsection{Pruning Function Definition}

To define the threshold $\tau$ for filtering, we calculate the average pixel intensity matrices $\bar{x}_0$ and $\bar{x}_1$ across all training samples in each binary class. Specifically, $\bar{x}_0(i, j)$ and $\bar{x}_1(i, j)$ represent the average intensity values at position $(i, j)$ for classes 0 and 1, respectively. Using these averages, we prune data by setting values to zero at positions $(i, j)$ where both class averages fall below the threshold $\tau$, as follows:
\begin{equation}
x_{\tau}(i, j) = 
\begin{cases} 
    0, & \text{if } \bar{x}_0(i,j) < \tau \text{ and } \bar{x}_1(i,j) < \tau, \\
    x(i,j), & \text{otherwise,} 
\end{cases}
\end{equation}
where $x(i,j)$ represents the original pixel intensity at position $(i, j)$. This operation generates a pruned dataset $\mathcal{X}_{\tau}$ that retains only grid positions with sufficient intensity to contribute effectively to classification.

To determine the optimal threshold $\tau^*$, we maximize the test accuracy $\text{Acc}_{\text{test}}(\mathcal{X}_{\tau})$:
\begin{equation}
\tau^* = \arg \max_{\tau \in [0, \tau_{\max}]} \text{Acc}_{\text{test}}(\mathcal{X}_{\tau}),
\end{equation}
where $\tau_{\max}$ is the upper bound on $\tau$. This optimization is equivalent to minimizing the negative accuracy $-\text{Acc}_{\text{test}}(\mathcal{X}_{\tau})$ over the interval $[0, \tau_{\max}]$.

\subsection{Gradient-Based Optimization via L-BFGS-B}

To solve the threshold optimization problem, we use the L-BFGS-B algorithm \cite{zhu1997algorithm}, which efficiently manages high-dimensional optimization with limited memory. The algorithm iteratively adjusts $\tau$ by minimizing the negative accuracy:
\begin{equation}
f(\tau) = -\text{Acc}_{\text{test}}(\mathcal{X}_{\tau}),
\end{equation}
updating $\tau$ at each step according to:
\begin{equation}
\tau_{k+1} = \tau_k - \alpha_k \, H_k \nabla f(\tau_k),
\end{equation}
where $\alpha_k$ is the step size, $H_k$ is an approximation of the inverse Hessian matrix, and $\nabla f(\tau_k)$ is the gradient of $f(\tau_k)$ with respect to $\tau$. The gradient $\nabla f(\tau_k)$ is computed as:

\begin{equation}
\nabla f(\tau_k) = \sum_{(i,j)} \frac{\partial f}{\partial x_{\tau}(i,j)} \cdot \frac{\partial x_{\tau}(i,j)}{\partial \tau}.
\end{equation}

This approach, as detailed in Algorithm~\ref{alg:thresh_opt}, leverages gradient information to efficiently converge on the optimal threshold $\tau^*$, ensuring a refined selection of relevant data regions.

\begin{algorithm}
\caption{Bi-Level Threshold Optimization for QNN}
\begin{algorithmic}[1]
\Require Binary class pair $(c_0, c_1)$, dataset $(x, y)$, encoding function $f_{\text{enc}}$, grid size $s$, epochs, threshold range
\Ensure Optimized threshold $\tau^*$ and classification accuracy

\State Filter and resize data for selected class pair $(c_0, c_1)$, converting labels to binary
\State Calculate average values for each class and set pixels to zero where averages are below $\tau$
\State Convert processed images to quantum circuits using $f_{\text{enc}}$

\State Initialize QNN model with angle encoding
\State Define bi-level optimization: 
\State \quad Inner level: apply threshold, filter data, train QNN on $(x_{\text{train}}, y_{\text{train}})$
\State \quad Outer level: maximize test accuracy on $(x_{\text{test}}, y_{\text{test}})$
\State Optimize over $\tau$ within range using L-BFGS-B
\State Calculate entanglement entropy of final model on test set
\State \Return optimal threshold $\tau^*$, test accuracy, and entanglement entropy
\end{algorithmic}
\label{alg:thresh_opt}
\end{algorithm}

\subsection{Constraint Handling and Convergence Criteria}

The box constraints $0 \leq \tau \leq \tau_{\max}$ are enforced within L-BFGS-B by projecting any out-of-bounds updates back to the feasible region, ensuring that the threshold remains within the desired range throughout the optimization. Convergence is achieved when the norm of the projected gradient $\|\nabla f(\tau_k)\|_{\infty}$ falls below a predefined tolerance, $\epsilon$, or when the maximum number of iterations is reached:
\begin{equation}
\|\nabla f(\tau_k)\|_{\infty} < \epsilon.
\end{equation}

Additionally, the accuracy is evaluated on a validation set $\mathcal{X}_{\text{val}}$ after each iteration to ensure that the optimization process generalizes effectively to unseen data, mitigating overfitting to the training set.

\subsection{Approximation of the Inverse Hessian}

To achieve computational efficiency, L-BFGS-B approximates the Hessian matrix rather than computing it directly \cite{zhu1997algorithm}. Using a limited history of $m$ gradient and position vectors $\{s_i, y_i\}_{i=k-m}^{k-1}$, where $s_i = \tau_{i+1} - \tau_i$ and $y_i = \nabla f(\tau_{i+1}) - \nabla f(\tau_i)$, the inverse Hessian approximation $H_k$ is updated using the recursive formula:
\begin{equation}
H_{k+1} = V_k^T H_k V_k + \rho_k s_k s_k^T,
\end{equation}
where $\rho_k = (y_k^T s_k)^{-1}$ and $V_k = I - \rho_k y_k s_k^T$. This recursive update captures the essential curvature information while minimizing memory overhead, making the optimization feasible for high-dimensional datasets.

After identifying the optimal threshold $\tau^*$, we generate the filtered datasets $\mathcal{X}_{\tau^*}$ for training and testing, where irrelevant positions have been pruned based on $\tau^*$. The QNN is then trained on $\mathcal{X}_{\tau^*}$, leveraging the optimized threshold to focus on critical data regions while minimizing computational load and enhancing classification performance. Test accuracy $\text{Acc}_{\text{test}}(\mathcal{X}_{\tau^*})$ is subsequently evaluated to confirm the effectiveness of threshold optimization.

\section{Experiments}

\subsection{Setup}
To evaluate our approach, we conducted binary classification experiments on multiple benchmark datasets: MNIST, FashionMNIST, CIFAR, and PneumoniaMNIST. These datasets were chosen to assess the model's adaptability across varied image types and complexities. The experiments were implemented using a compact three-layer QNN model, which serves as the base architecture for all encoding techniques tested. The QNN model applies a parameterized quantum circuit built with data qubits and a readout qubit, initialized with an $X$ gate followed by a Hadamard gate to prepare it in superposition. The circuit incorporates both entangling $\text{XX}$ and $\text{ZZ}$ gates between the data qubits and the readout qubit, with learnable parameters to capture complex dependencies within the input data. A final Hadamard gate is applied to the readout qubit before measurement, completing the encoding and entanglement process for each layer. 

\subsubsection{Encoding and Preprocessing Techniques}
For encoding, Angle and Amplitude methods were applied directly to the data. In contrast, ATP, SQE, and PCA were used beforehand, each applying a different approach to prepare the data before encoding. Specifically, ATP and PCA refine data structure before encoding, while SQE focuses on single-qubit encoding. The results presented here are from experiments where these preparatory methods were followed by Angle encoding. Additional experiments with direct Amplitude encoding, as well as further analyses, are included in the supplementary materials and ablation study section.

\subsubsection{Realistic Noise Conditions}
To evaluate robustness in realistic settings, depolarizing noise was introduced at intensities of 3\%, 5\%, and 10\%, allowing for a comparison of baseline performance on clean data and the model’s resilience under these quantum noise levels. This setup enabled a thorough assessment of each encoding method’s effectiveness, revealing how the QNN model performs across both ideal and noisy environments.

\subsection{Manual Threshold Pruning}

\begin{figure}[h]
\centering
\includegraphics[width=\columnwidth]{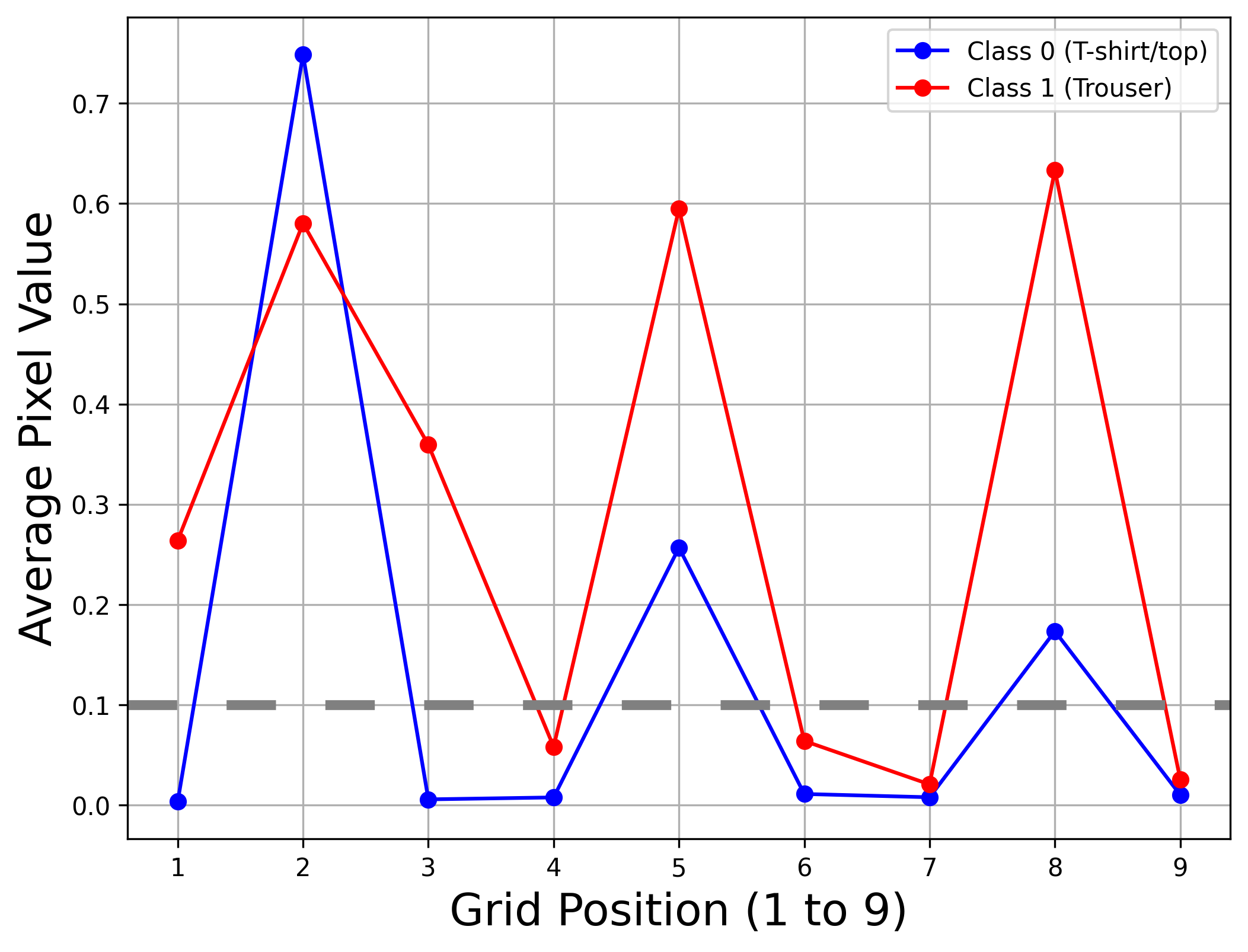}
\caption{Data distribution across positions in FashionMNIST for two classes (T-shirt/top and Trouser), with dashed lines marking threshold levels. Positions with lower variance are pruned to streamline training and focus on more informative regions.}
\label{fig:pixel_intensity_thresh}
\end{figure}

In this part, we examine the effect of pruning with different thresholds on QNN performance. Figure~\ref{fig:pixel_intensity_thresh} illustrates the data distribution for each position within a 3x3 division of the image, where the values represent average intensities across samples from a specific class. As shown, positions with lower information content tend to have closely overlapping values between classes, which can make training more challenging by introducing less distinctive information into the QNN.

\begin{figure}[h]
\centering
\includegraphics[width=\columnwidth]{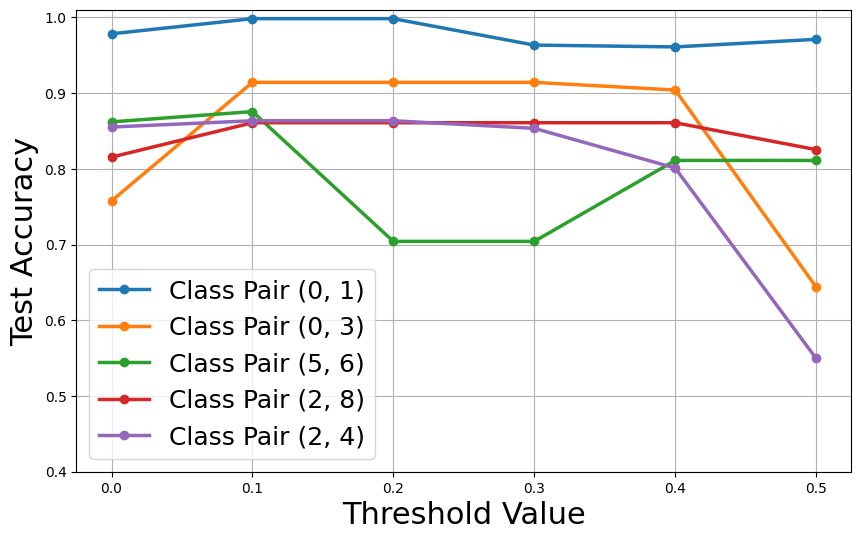}
\caption{Test accuracy for MNIST class pairs with varying pruning thresholds. Moderate thresholds improve accuracy, while higher thresholds may exclude key information.}
\label{fig:mnist_testacc}
\end{figure}

\begin{figure}[h]
\centering
\includegraphics[width=\columnwidth]{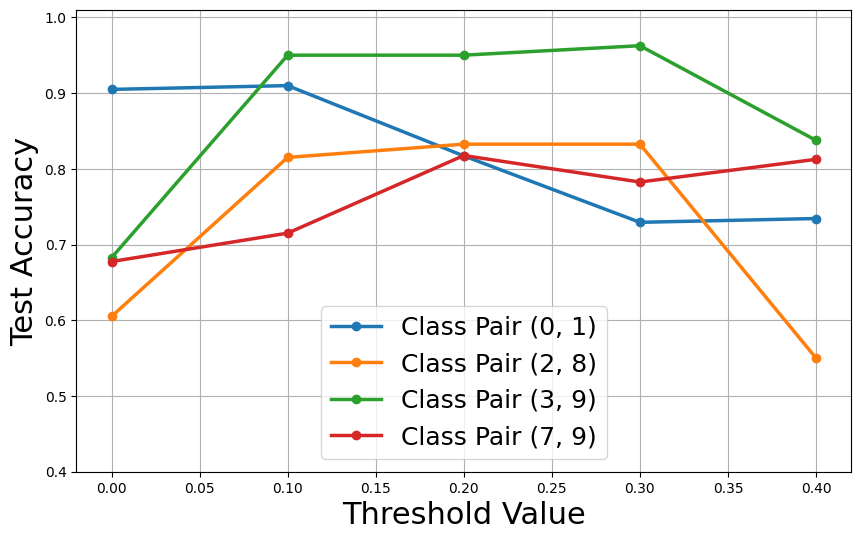}
\caption{Test accuracy for FashionMNIST class pairs with varying thresholds. Similar to MNIST, a threshold around 0.3 generally provides optimal performance.}
\label{fig:fashion_testacc}
\end{figure}

To further investigate the impact of these thresholds on performance, we apply a range of manual thresholds across different class pairs in both MNIST and FashionMNIST datasets. Figures~\ref{fig:mnist_testacc} and~\ref{fig:fashion_testacc} show that moderate pruning generally leads to improved accuracy by eliminating non-essential features while preserving critical distinctions. In Figure~\ref{fig:mnist_testacc}, results for MNIST indicate that moderate thresholds enhance accuracy, while higher thresholds reduce it by removing valuable details. Similarly, for FashionMNIST (Figure~\ref{fig:fashion_testacc}), the optimal thresholds cluster between 0.1 and 0.3, but specific values vary with data distribution characteristics.

These results demonstrate that the optimal pruning level is influenced by the data distribution, motivating the need for an adaptive thresholding approach. The following section details the results from the bi-level optimization method used in our framework to automate threshold selection based on each dataset’s characteristics.

\subsection{Performance Results}
Tables \ref{table:accuracy-performance} and \ref{table:entropy-performance} present the classification accuracy and EE, respectively, for different encoding techniques across the four datasets used in the QNN binary classification tasks. For MNIST and FashionMNIST, multiple class pairs were chosen to evaluate the model’s capability in distinguishing various subsets, providing a broader assessment of encoding effectiveness. Across the majority of datasets and class pairs, ATP achieved the highest accuracy. ATP also consistently minimized EE compared to other methods, indicating a more efficient use of quantum resources by lowering complexity without compromising accuracy.

\begin{table}[!t]
\centering
\caption{Performance of encoding techniques on various classes for binary classifications (Accuracy).}
\label{table:accuracy-performance}
\begin{tabular}{l|c|c|c|c|c}
    Classes & Angle & Amplitude & ATP & PCA & SQE \\
    \hline
    \multicolumn{6}{c}{MNIST} \\
    \hline
    (0,1) & 96.0 & 95.5 & \textbf{99.0} & \textbf{99.0} & 88.0 \\
    (0,3) & 89.0 & 88.5 & \textbf{91.0} & 88.0 & 86.0 \\
    (2,4) & 85.0 & 84.0 & \textbf{86.0} & 84.5 & 82.0 \\
    (5,6) & 86.0 & 85.5 & \textbf{87.0} & 85.0 & 83.5 \\
    (2,8) & 81.0 & 79.5 & 83.0 & \textbf{86.0} & 78.5 \\
    \hline
    \multicolumn{6}{c}{Fashion MNIST} \\
    \hline
    (0,1) & 88.5& 88.0 & \textbf{91.5} & 88.5 & 86.0 \\
    (2,8) & \textbf{86.0} & 84.5 & \textbf{86.0} & \textbf{86.0} & 83.0 \\
    (3,9) & \textbf{94.0} & 87.0 & \textbf{94.0} & 93.0 & 91.0 \\
    (7,9) & 82.0 & 78.0 & \textbf{83.0} & 79.0 & 77.0 \\
    \hline
    \multicolumn{6}{c}{CIFAR} \\
    \hline
    (0,1) & 70.0 & 68.5 & \textbf{74.2} & 68.0 & 66.0 \\
    \hline
    \multicolumn{6}{c}{PneumoniaMNIST} \\
    \hline
    (0,1) & 81.0& 68.5 & \textbf{87.0} & 80.0 & 75.5 \\
\end{tabular}
\end{table}

\begin{table}[!t]
\centering
\caption{Entanglement entropy of encoding techniques on various classes for binary classifications.}
\label{table:entropy-performance}
\begin{tabular}{l|c|c|c|c|c}
    Classes & Angle & Amplitude & ATP & PCA & SQE \\
    \hline
    \multicolumn{6}{c}{MNIST} \\
    \hline
    (0,1) & 0.67 & 0.52 & \textbf{0.39} & 0.60 & 0.45 \\
    (0,3) & 0.59 & 0.53 & \textbf{0.34} & 0.55 & 0.44 \\
    (2,4) & 0.77 & 0.64 & 0.45 & 0.56 & \textbf{0.41} \\
    (5,6) & 0.82 & 0.53 & 0.41 & 0.61 & \textbf{0.39} \\
    (2,8) & 0.63 & 0.56 & \textbf{0.34} & 0.36 & \textbf{0.34} \\
    \hline
    \multicolumn{6}{c}{Fashion MNIST} \\
    \hline
    (0,1) & 0.56 & 0.47 & \textbf{0.35} & 0.58 & 0.39 \\
    (2,8) & \textbf{0.52} & 0.50 & 0.35 & 0.41 & 0.38 \\
    (3,9) & 0.54 & 0.62 & \textbf{0.38} & 0.55 & 0.42 \\
    (7,9) & 0.59 & 0.58 & 0.41 & 0.64 & \textbf{0.43} \\
    \hline
    \multicolumn{6}{c}{CIFAR} \\
    \hline
    (0,1) & 0.63 & 0.51 & 0.46 & 0.65 & \textbf{0.43} \\
    \hline
    \multicolumn{6}{c}{PneumoniaMNIST} \\
    \hline
    (0,1) & 0.88 & 0.79 & \textbf{0.37} & 0.59 & 0.42 \\
\end{tabular}
\end{table}

To assess the encoding methods' resilience to noise, we examined model performance under various levels of depolarizing noise (3-10\%). Table~\ref{table:accuracy-noise-performance} highlights that encoding techniques with lower entanglement entropy, such as ATP and SQE, showed stronger robustness with accuracy reductions typically limited to 3–8 points. In contrast, Angle, Amplitude, and PCA experienced more substantial drops in accuracy of 4–17 points. These results emphasize ATP and SQE's effectiveness in preserving model accuracy under challenging noise conditions, establishing a baseline for QNN stability in practical, noisy environments.

\begin{table}[!t]
\centering
\caption{Performance of encoding techniques on various classes for binary classifications under depolarizing noise (Accuracy).}
\label{table:accuracy-noise-performance}
\begin{tabular}{l|c|c|c|c|c}
    Classes & Angle & Amplitude & ATP & PCA & SQE \\
    \hline
    \multicolumn{6}{c}{MNIST} \\
    \hline
    (0,1) & 87.0 & 86.4 & \textbf{89.0} & \textbf{89.0} & 88.2 \\
    (0,3) & 82.4 & 80.5 & \textbf{83.0} & 78.9 & 81.9 \\
    (2,4) & \textbf{80.2} & 68.8 & 71.0 & 79.7 & 79.8 \\
    (5,6) & 70.8 & 69.4 & 73.0 & 68.5 & \textbf{74.0} \\
    (2,8) & 68.8 & 57.4 & \textbf{76.5} & 74.9 & 75.2 \\
    \hline
    \multicolumn{6}{c}{Fashion MNIST} \\
    \hline
    (0,1) & 80.5 & 79.0 & \textbf{84.9} & \textbf{84.9} & 82.1 \\
    (2,8) & 76.1 & 75.4 & 79.3 & 78.6 & \textbf{79.5} \\
    (3,9) & 84.3 & 78.2 & \textbf{87.9} & 86.5 & 86.7 \\
    (7,9) & 70.0 & 66.2 & \textbf{68.4} & 73.2 & 74.5 \\
    \hline
    \multicolumn{6}{c}{CIFAR} \\
    \hline
    (0,1) & 60.4 & 59.0 & \textbf{61.0} & 57.8 & 58.5 \\
    \hline
    \multicolumn{6}{c}{PneumoniaMNIST} \\
    \hline
    (0,1) & 70.5 & 63.4 & \textbf{76.0} & 67.3 & 68.7 \\
\end{tabular}
\end{table}

\subsection{Adversarial Robustness}

To evaluate the resilience of each encoding method against adversarial attacks, we applied Fast Gradient Sign Method (FGSM) with an attack strength of $\epsilon = 0.3$ across the datasets. Table~\ref{table:adversarial-performance} provides accuracy results for binary classification tasks under these adversarial conditions. In most cases, direct encoding methods like Angle and Amplitude encoding were generally more susceptible to the attacks, displaying substantial accuracy reductions. ATP, PCA, and SQE exhibited a moderate level of robustness, although performance varied across datasets. Notably, ATP and PCA maintained comparable accuracy levels in several cases, but in certain tasks (PneumoniaMNIST and CIFAR), Angle encoding unexpectedly achieved the highest accuracy, suggesting that additional adversarial training methods may be required to further improve model resilience.

\begin{table}[h]
\centering
\caption{Performance of encoding techniques on various classes for binary classifications under FGSM attack with $\epsilon = 0.3$ (Accuracy).}
\label{table:adversarial-performance}
\begin{tabular}{l|c|c|c|c|c}
    Classes & Angle & Amplitude & ATP & PCA & SQE \\
    \hline
    \multicolumn{6}{c}{MNIST} \\
    \hline
    (0,1) & 62.0 & 58.4 & \textbf{67.2} & 66.5 & 65.0 \\
    (0,3) & 63.2 & 61.0 & 67.3 & \textbf{69.0} & 63.7 \\
    (2,4) & 55.2 & 54.7 & 60.0 & \textbf{61.5} & 59.0 \\
    (5,6) & 64.3 & 61.5 & \textbf{66.7} & 65.0 & 62.5 \\
    (2,8) & 48.6 & 50.5 & 56.0 & \textbf{57.8} & 54.0 \\
    \hline
    \multicolumn{6}{c}{Fashion MNIST} \\
    \hline
    (0,1) & 68.4 & 66.0 & 72.5 & \textbf{73.0} & 70.2 \\
    (2,8) & 66.2 & 65.5 & 70.3 & \textbf{71.0} & 69.0 \\
    (3,9) & \textbf{77.5} & 74.2 & 76.0 & 75.8 & 73.5 \\
    (7,9) & 60.0 & 59.2 & \textbf{64.8} & 63.5 & 62.3 \\
    \hline
    \multicolumn{6}{c}{CIFAR} \\
    \hline
    (0,1) & 55.0 & 51.4 & 59.5 & \textbf{61.2} & 58.0 \\
    \hline
    \multicolumn{6}{c}{PneumoniaMNIST} \\
    \hline
    (0,1) & \textbf{66.5} & 62.0 & 64.0 & 65.2 & 63.4 \\
\end{tabular}
\end{table}

\begin{figure}[h]
\centering
\includegraphics[width=\columnwidth]{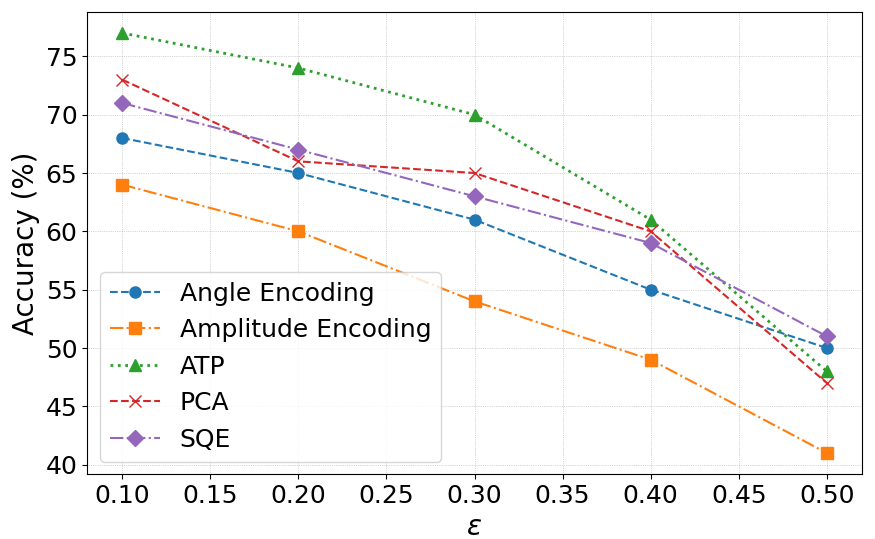}
\caption{Effectiveness of different encoding methods against adversarial attacks after adversarial training, evaluated at varying attack strengths (measured by $\epsilon$) for (0,1) classification tasks on MNIST. ATP shows higher robustness across most $\epsilon$ values, though all models experience performance declines as $\epsilon$ increases.}
\label{fig:robustness}
\end{figure}

The encoding techniques used in this study, including ATP, lack inherent robustness against adversarial attacks, as shown in prior evaluations. To address this, adversarial training was applied as an additional defense. Figure~\ref{fig:robustness} presents post-training accuracy for each encoding method under FGSM attacks, highlighting ATP and PCA’s enhanced performance due to their filtering of non-informative features, which decreases sensitivity to irrelevant perturbations. Additional results for other attack models are provided in the supplementary material.

As shown in Figure~\ref{fig:robustness}, adversarial training enhances resilience across most encoding methods, particularly at lower $\epsilon$ values. ATP consistently achieves higher accuracy across attack strengths, maintaining a moderate lead over PCA and SQE despite performance declines as $\epsilon$ increases. Amplitude and Angle encoding remain more susceptible to attacks, especially at higher $\epsilon$ values, indicating lower robustness. These results underscore ATP’s advantage in accuracy preservation, though further measures may be needed for robustness under stronger adversarial conditions.

\subsection{Experiments on IBM Quantum Hardware}
To evaluate ATP’s robustness on real quantum hardware, we extended our experiments to the IBM Sherbrooke backend using Qiskit’s Runtime Service. We implemented the proposed framework with Qiskit’s native \texttt{EstimatorQNN} interface, using default settings for error mitigation and backend resilience. The model was trained with a COBYLA optimizer for 30 iterations. Across five MNIST class subsets, ATP demonstrated an average performance improvement of 7\% compared to the direct encoding approach, with notable accuracy gains on the (2,4) and (2,8) classification tasks, the latter showing the highest improvement.

\section{Discussion and Limitations}
While ATP achieved notable performance gains across datasets, some limitations remain. First, we focus on binary classification for two main reasons: comparability with existing literature and the efficiency of quantum measurement processes. Comparability is important, as most prior works predominantly use binary classification to evaluate state-of-the-art QNN implementations. Moreover, binary classification simplifies quantum measurement, as the class label can be inferred from a single-qubit von Neumann measurement—such as the sign of the expectation value of the \( \sigma_z \) observable. This approach, including measuring only the last qubit to determine the output label with minimal circuit complexity, has been well-documented in prior studies \cite{hur2022quantum,wu2023quantumdarts}. In contrast, most multi-class classification methods require a hybrid between classical and quantum models \cite{bokhan2022multiclass}, making binary classification the preferred choice both in the literature and in our study. Additionally, ATP’s optimization process incurs moderate computational overhead, with threshold selection and pruning taking approximately 15\% longer than direct angle encoding (compared to 12\% for PCA and 4\% for SQE). While this may affect efficiency in larger applications, further optimizations in the thresholding algorithm could help reduce these demands.

Future work could also explore ATP's adaptability to datasets with varied underlying distributions, as current evaluations focus on standard QNN benchmarks. Investigating performance on more diverse datasets may offer insights for improving ATP’s robustness in broader real-world scenarios.

\section{Conclusion}
We presented Adaptive Threshold Pruning (ATP) to address the resource constraints of quantum systems by removing non-essential input features before encoding, enabling more efficient use of qubits. Rather than reducing circuit complexity directly, ATP lowers input redundancy, which leads to reduced entanglement and improved computational efficiency. Across multiple datasets, ATP consistently outperforms other encoding methods by achieving higher accuracy with lower entanglement entropy. By adapting pruning thresholds to the variance structure of the data, ATP offers a flexible and scalable encoding strategy that enhances QNN performance in resource-limited settings.

{
    \small
    \bibliographystyle{ieeenat_fullname}
    \bibliography{main}
}

\end{document}